\documentclass{article}
\usepackage{amsfonts,amssymb}
\title{Two-dimensional algebro-geometric difference operators}
\author{A.~A.~Oblomkov%
\thanks{Department of Mathematics and Mechanics,
Moscow State University, Moscow, 119899, Russia.}
\thanks{Independent University of Moscow, 
Bolshoy Vlasyevskiy per. 11, Moscow, 121002, Russia. 
{\tt e-mail: oblomkov@mccme.ru}}, 
A.~V.~Penskoi%
\thanks{Centre de recherches math\'ematiques, Universit\'e de Montr\'eal,
C.~P.~6128, succ. Centre-ville, Montr\'eal, 
Qu\'ebec, H3C 3J7, Canada. {\tt e-mail: penskoi@crm.umontreal.ca}}}
\date{}
\begin{document}
\maketitle
\abstract{A generalized inverse problem for a
two-dimensional difference operator is introduced.
A new construction of the algebro-geometric difference operators
of two types first considered by I.M.Krichever and S.P.Novikov is
proposed.}

\section{Introduction}

The notion of a finite-gap with respect to one
energy level Schr\"odinger operator was introduced
by B.~A.~Dubrovin, I.~M.~Krichever and S.~P.~Novikov
in the paper~\cite{DKN}.
The paper of S.~P.~Novikov and A.~P.~Veselov~\cite{NV}
deals with some class of two-dimensional Schr\"odinger
operators called potential operators. In this
paper S.~P.~Novikov and A.~P.~Veselov solved the inverse
scattering problem.
In the paper~\cite{K},
I.~M.~Krichever has introduced
a similar theory for difference operators.
The recent papers~\cite{N,NVC,ND}
deal with different natural generalizations of two-dimensional
difference operators defined on regular graphs and lattices.
In particular in the paper \cite{N} (see also appendix I in the
paper \cite{NVC}) in the context of the discrete Laplace
transformations S.~P.~Novikov introduced an important class of
the difference operators on equilateral triangular lattice.
These papers stimulated new research in this area (see
the review ~\cite{ND}).

In the present paper we propose a generalized
inverse problem and a new construction of two-dimensional
algebro-geometric operators both in Krichever's and Novikov's
classes.

Let $L$ be a two-dimensional difference
operator (of order $2K$)
\begin{equation}\label{KL}
(L\psi)_{nm}=\sum\limits_{i,j,|i|\le K,|j|\le K}%
a_{nm}^{ij}\psi_{n+i,m+j}
\end{equation}
with periodic coefficients
$$
a^{ij}_{n+N,m}=a^{ij}_{n,m+M}=a^{ij}_{nm}.
$$
Consider a space of Floquet 
functions
$$
\psi_{n+N,m}=w_1\psi_{n,m},\quad
\psi_{n,m+M}=w_2\psi_{n,m}.
$$
This space is finite-dimensional and
the operator $L$ induces in this space a linear
operator $L(w_1,w_2).$
The characteristic equation of this operator
$$
Q(w_1,w_2,E)=\det(E\cdot\mbox{Id}-L(w_1,w_2))=0
$$
defines a two-dimensional algebraic
variety $M^2.$ 
A point of $M^2$ corresponds to a unique
eigenvector $\psi_{nm}$ of the operator $L$
$$
(L\psi)_{nm}=E\psi_{nm}
$$
such that $\psi_{00}=1$. All other components
$\psi_{nm}$ are meromorphic functions on $M^2.$
Consider a curve $\Gamma\subset M^2$
corresponding to the ``zero-energy level''
$$
\Gamma=\{w_1,w_2|Q(w_1,w_2,0)=0\}.
$$
The functions $\psi_{nm}$ are meromorphic
on $\Gamma$.

We can consider the two following problems.
\par\noindent 1) The direct spectral problem. Find explicitly
the ``spectral data'' of the operator $L$
(i.e. a set of geometric data like a curve $\Gamma,$ divisors of poles
of $\psi_{nm}$ etc.) which determines the operator $L$ uniquely.
\par\noindent 2) The inverse spectral problem. Find explicitly
the operator $L$ using the ``spectral data''.

Both problems are complicated. 
It is nearly imposible to solve either of them in a general case.
We can, however, consider
a generalized inverse problem which 
consists of finding 
a set of geometric data with the following properties:
\par\noindent 1) The set of geometric data defines uniquely
a family of functions $\psi_{nm}$ defined on an algebraic
complex curve $\Gamma,$ 
\par\noindent 2) These functions satisfy the equation 
$L\psi=0$ for some operator $L$ of the form~(\ref{KL}), 
\par\noindent 3) The operator
$L$ is uniquely defined by the equation $L\psi=0$ and the coefficients
$a_{nm}^{ij}$ can be found explicitly.

This problem is solved for
some particular operators
in the paper~\cite{K}. I.~M.~Krichever calls such operators ``integrable''
but we will use the term ``algebro-geometric''.

Our goal is to find algebro-geometric operators. We found two examples
which can be of interest. 

The first example is provided by 
operators of the form
\begin{equation}\label{L}
(L\psi)_{nm}=a_{nm}\psi_{n-1,m}+b_{nm}\psi_{n+1,m}+c_{nm}\psi_{n,m-1}%
+d_{nm}\psi_{n,m+1}+v_{nm}\psi_{nm}.
\end{equation}
A value of $(L\psi)_{nm}$ depends only on values of $\psi$ at 
the points 
$$
(n-1,m), (n+1,m), (n,m-1), (n,m+1), (n,m)
$$ 
which form
a cross in the plane $(n,m).$ 
We will call such an operator
``cross-shaped''.
These operators were considered by I.~M.~Krichever
in~\cite{K}. Algebro-geometric operators of the form~(\ref{L})
found by I.~M.~Krichever correspond to 
a curve $\Gamma'\subset M^2$ whose image under the
projection on the $E$-plane is the whole $E$-plane. 
The corresponding problem is $L\psi=E\psi,$ where
both $E$ and $\psi$  are functions defined on $\Gamma'.$
In the present paper we deal with a different type of 
algebro-geometric operators of the form~(\ref{L}) which
corresponds to the ``zero energy level'' curve.
The corresponding problem is $L\psi=0.$

The other example is more complicated and maybe
more interesting.
Consider a triangular lattice in a plane.
We will use as coordinates triples of integers $k,l,m$ such that
$k+l+m=0.$ On such a lattice we can consider an operator of the form
\begin{equation}\label{L1}
(L\psi)_{klm}=a_{klm}\psi_{k,l+1,m-1}+b_{klm}\psi_{k,l-1,m+1}+%
c_{klm}\psi_{k+1,l-1,m}+
\end{equation}
$$
+d_{klm}\psi_{k-1,l+1,m}+%
f_{klm}\psi_{k+1,l,m-1}+g_{klm}\psi_{k-1,l,m+1}.
$$
A value of $(L\psi)_{klm}$ depends only on values of $\psi$ at
the points 
$$
(k,l+1,m-1), (k,l-1,m+1), (k+1,l-1,m), (k-1,l+1,m),
$$
$$
(k+1,l,m-1), (k-1,l,m+1)
$$ 
which form
a hexagon in the plane $k,l,m.$
We will call such an operator ``hexagonal''.
In this case our lattice is not rectangular,
nevertheless we can consider the generalized 
inverse problem and solve it.

As we have already mentioned, this class of the operators
has been introduced in the context of the discrete Laplace transformation
by S.~P.~Novikov~\cite{N,NVC,ND}.

It should be remarked that our formulas in the
sections~\ref{cross},~\ref{hexagonal} are not unique. We can
choose other singularity structures for $\psi$-function
(for example using a rotation of the plane $(n,m)$ by $\frac{\pi}{2}$
in the case of operators of the form~(\ref{L})) and obtain other 
algebro-geometric operators.

\section{Notation and conventions}

We use the notations and conventions of paper~\cite{D}.
In particular, our conventions are the following.
A basis of cycles $a_1\dots,a_g,b_1,\dots,b_g$ is chosen 
in such a way that
$$
a_i\circ a_j=b_i\circ b_j=0,\quad a_i\circ b_j=\delta_{ij},
\quad i,j=1,\dots,g,
$$
where $g$ is the genus of a non-singular curve $\Gamma.$
A basis of holomorphic differentials $\omega_1,\dots,\omega_g$
is choosen in such a way that
$$
\oint_{a_j}\omega_k=2\pi i\delta_{jk},\quad j,k=1,\dots,g.
$$
We define the Jacobian $J(\Gamma)$ as
$\mathbb{C}^g/\{2\pi i M+BN\},$
where $M,N\in\mathbb{Z}^g,$
$B$ is a matrix of $b$-periods of $\omega_i$
$$
B_{jk}=\oint_{b_j}\omega_k,\quad j,k=1,\dots,g.
$$
We denote by $\Omega_{PQ}$ the Abel differential of the third
kind, i.e. a differential with unique poles
at the points $P$ and $Q$ and residues $+1$ and $-1$ at these points
respectively, we denote by $U_{PQ}$ the vector of $b$-periods
of $\Omega_{PQ},$ we denote by
$\mathcal{K}$ the vector of the
Riemann constants.

We define the $\Theta$-function as
$$
\Theta(z)=\sum_{N\in{\mathbb Z}^g}\exp%
\left(\frac{1}{2}\langle BN,N\rangle+\langle N,z\rangle\right),
$$
where $z=(z_1,\dots,z_g)\in{\mathbb C}^g$ and
$\langle\,,\rangle$ is a euclidean scalar product 
$\langle x,y\rangle=\sum_{i=1}^gx_iy_j.$

We use the following natural convention: if $n$ is a negative integer
then a zero (pole) of the $n$th order is a pole (zero) of the $|n|$th order.

\section{The cross-shaped operators: Krichever's class}\label{cross}

Consider an arbitrary
two-dimensional difference operator $L$ of the form~(\ref{L}).
Our goal is to find some solution of the generalized inverse problem
stated in the introduction.

Our construction is as follows. Let $\Gamma$ be a nonsingular curve of genus
$g.$ Let $P_i^{\pm},i=1,2,3,$ be six points on $\Gamma.$ 
Let $\mathcal D$ be a generic divisor of the form
$\mathcal{D}=P_1+\dots+P_g$ 
such that the points
$P_k$ are different from the $P_i^{\pm}.$
Consider a function
$\phi_{\alpha\beta\gamma},\alpha,\beta,\gamma\in\mathbb{Z},$ defined
on $\Gamma$ such that:

\noindent 1) If a point $P\in\Gamma\setminus\{P_1^{\pm},P_2^{\pm},P_3^{\pm}\}$
is a pole of $\phi_{\alpha\beta\gamma},$ then $P$ is one of the points
$P_k;$

\noindent 2) The function $\phi_{\alpha\beta\gamma}$ has a zero of $\alpha$th order
in $P_1^+$ and a pole of $\alpha$th order in $P_1^-,$ the same structure for 
$\beta$ and $P_2^{\pm}$, $\gamma$ and $P_3^{\pm}.$

\noindent{\sc Lemma} 1) Such a function $\phi_{\alpha\beta\gamma}$ exists and
is unique up to multiplication by a constant.

\noindent 2) The explicit formula for $\phi_{\alpha\beta\gamma}$ is:
$$
%\psi_{\alpha\beta\gamma}(P)=
r_{\alpha\beta\gamma}\cdot%
\exp\int_{P_0}^{P}\!\!\!(\alpha\Omega_1+\beta\Omega_2+\gamma\Omega_3)\cdot%
\frac{\Theta(A(P)+\alpha U_1+\beta U_2+\gamma U_3-A(\mathcal{D})-\mathcal{K})}%
{\Theta(A(P)-A(\mathcal{D})-\mathcal{K})},
$$
where $r_{\alpha\beta\gamma}$ is an arbitrary constant, $P_0$ is 
a fixed point defining the Abel transform $A$ (it should be remarked
that the paths of integration in $\int_{P_0}^P$ and in the Abel transform
are the same), $\Omega_i=\Omega_{P_i^+P_i^-}$, $U_i=U_{P_i^+P_i^-}.$

\noindent{\sc Proof} is done by a standard reasoning of the
theory of the algebro-geometric integration. $\square$

The key idea of the construction of our functions
$\psi_{nm}$ is a convinient relabelling in following way: 
$\psi_{nm}=\phi_{\alpha\beta\gamma},$
where
$$
\left\{%
\begin{array}{l}
\alpha(n,m)=\frac{\displaystyle 2-n-m}{\displaystyle 2},
\beta(n,m)=\frac{\displaystyle n-m}{\displaystyle 2}, 
\gamma(n,m)=\frac{\displaystyle n-m}{\displaystyle 2}, \\
\mbox{if} \quad n+m=0 \pmod 2,\\
\alpha(n,m)=\frac{\displaystyle 3-n-m}{\displaystyle 2}, 
\beta(n,m)=\frac{\displaystyle -1+n-m}{\displaystyle 2},
\gamma(n,m)=\frac{\displaystyle 1+n-m}{\displaystyle 2},\\ 
\mbox{if} \quad n+m=1 \pmod 2.
\end{array}
\right.
$$

We will use a vectorial notation for the triples,
i.~e. the representation
of a triple $\alpha,\beta,\gamma$ as a vector 
$\alpha\mathbf{i}+\beta\mathbf{j}+\gamma\bf{k}.$
For example, we will sometime write
$\phi_{\alpha\mathbf{i}+\beta\mathbf{j}+\gamma\bf{k}}$
instead of $\phi_{\alpha,\beta,\gamma}.$ This is useful because
for exemple, if $\mathbf{v}=\alpha\mathbf{i}+\beta\mathbf{j}+\gamma\bf{k},$
then we can write $\phi_{\mathbf{v}+\mathbf{i}}$ instead of
$\phi_{\alpha+1,\beta,\gamma}.$

We write $\psi_{nm}=\phi_{\mathbf{v}(n,m)},$
where $\mathbf{v}(n,m)=\alpha(n,m)\mathbf{i}+\beta(n,m)\mathbf{j}%
+\gamma(n,m)\mathbf{k},$ i.~e.
$$
\left\{%
\begin{array}{lll}
\mathbf{v}(n,m)=\frac{2-n-m}{2}\mathbf{i}+\frac{n-m}{2}\mathbf{j}+%
\frac{n-m}{2}\mathbf{k},& 
\mbox{if}& n+m=0 \pmod 2,\\
\mathbf{v}(n,m)=\frac{3-n-m}{2}\mathbf{i}+\frac{-1+n-m}{2}\mathbf{j}+%
\frac{1+n-m}{2}\mathbf{k},&
\mbox{if}& n+m=1 \pmod 2.
\end{array}
\right.
$$

We will also use the following notation
$$
\Theta(P,\alpha\mathbf{i}+\beta\mathbf{j}+\gamma\mathbf{k})%
=\Theta(P,\alpha,\beta,\gamma)=\Theta(A(P)+\alpha U_1+\beta U_2+%
\gamma U_3-A(\mathcal{D})-\mathcal{K}).
$$

Let us formulate our theorem.

\noindent{\sc Theorem 1.} Let a family $\psi_{mn}$ be defined
as stated above. Then
$L\psi=0$ if and only if 
the coefficients
$a_{nm}$, $b_{nm}$, $c_{nm}$, $d_{nm}$, $v_{nm}$ of the operator $L$
are defined up to a multiplication by a constant by the following
formulas:

\noindent 1) if $n+m\equiv 0 \pmod 2,$ then
$$
\frac{\displaystyle a_{nm}}{\displaystyle d_{nm}}=%
-\frac{\displaystyle r_{\mathbf{v}-\mathbf{j}}}%
{\displaystyle r_{\mathbf{v}+\mathbf{i}-\mathbf{j}}}\cdot%
\frac{\displaystyle \Theta(P^+_2,\mathbf{v}-\mathbf{j})}%
{\displaystyle \Theta(P^+_2,\mathbf{v}+\mathbf{i}-\mathbf{j})}\cdot%
\exp(-\!\!\int\limits^{P^+_2}_{P_0}\!\Omega_1),
$$
$$
\frac{\displaystyle b_{nm}}{\displaystyle d_{nm}}=-%
\frac{\displaystyle r_{\mathbf{v}-\mathbf{j}}}%
{\displaystyle r_{\mathbf{v}+\mathbf{k}}}\cdot%
\frac{\displaystyle \Theta(P^+_2,\mathbf{v}-\mathbf{j})%
\Theta(P^-_1,\mathbf{v}+\mathbf{i}-\mathbf{j})%
\Theta(P^-_3,\mathbf{v}+\mathbf{i}+\mathbf{k})}%
{\displaystyle \Theta(P^+_2,\mathbf{v}+\mathbf{i}-\mathbf{j})%
\Theta(P^-_1,\mathbf{v}+\mathbf{i}+\mathbf{k})%
\Theta(P^-_3,\mathbf{v}+\mathbf{k})}\times
$$
$$
\times\exp(\int\limits^{P^-_3}_{P_0}\!\Omega_1-%
\!\!\int\limits^{P^+_2}_{P_0}\!\Omega_1-%
\!\!\int\limits^{P^-_1}_{P_0}\!(\Omega_2+\Omega_3)),
$$
$$
\frac{\displaystyle c_{nm}}{\displaystyle d_{nm}}=%
\frac{\displaystyle r_{\mathbf{v}-\mathbf{j}}}%
{\displaystyle r_{\mathbf{v}+\mathbf{i}+\mathbf{k}}}\cdot%
\frac{\displaystyle \Theta(P^+_2,\mathbf{v}-\mathbf{j})%
\Theta(P^-_1,\mathbf{v}+\mathbf{i}-\mathbf{j})}%
{\displaystyle \Theta(P^+_2,\mathbf{v}+\mathbf{i}-\mathbf{j})%
\Theta(P^-_1,\mathbf{v}+\mathbf{i}+\mathbf{k})}\cdot%
\exp(-\!\!\int\limits^{P^+_2}_{P_0}\!\Omega_1-
\!\!\int\limits^{P^-_1}_{P_0}\!(\Omega_2+\Omega_3)),
$$
$$
\frac{\displaystyle v_{nm}}{\displaystyle d_{nm}}=%
\frac{\displaystyle r_{\mathbf{v}-\mathbf{j}}}%
{\displaystyle r_{\mathbf{v}}}\cdot%
\frac{\displaystyle \Theta(P_2^+,\mathbf{v}-\mathbf{j})%
\Theta(P_1^-,\mathbf{v}+\mathbf{i}-\mathbf{j})}%
{\displaystyle \Theta(P_2^+,\mathbf{v}+\mathbf{i}-\mathbf{j})%
\Theta(P_1^-,\mathbf{v}+\mathbf{i}+\mathbf{k})}\times
$$
$$
\times\exp(-\!\!\int\limits_{P_0}^{P_2^+}\Omega_1-%
\!\!\int\limits_{P_0}^{P_1^-}(\Omega_2+\Omega_3)+%
\!\!\int\limits_{P_0}^{P_2^-}\Omega_3)\cdot%
\biggl[\frac{\displaystyle\Theta(P_3^-,\mathbf{v}+\mathbf{i}+\mathbf{k})%
\Theta(P_2^-,\mathbf{v}+\mathbf{k})}%
{\displaystyle\Theta(P_3^-,\mathbf{v}+\mathbf{k})%
\Theta(P_2^-,\mathbf{v})}\times
$$
$$
\times\exp(\int\limits_{P_0}^{P_3^-}\Omega_1)-%
\frac{\displaystyle\Theta(P_2^-,\mathbf{v}+\mathbf{i}+\mathbf{k})}%
{\displaystyle\Theta(P_2^-,\mathbf{v})}\cdot%
\exp(\int\limits_{P_0}^{P_2^-}\Omega_1)\biggr],
$$
where $\mathbf{v}=\mathbf{v}(n,m),$

\noindent 2) if $n+m\equiv 1 \pmod 2,$ then
$$
\frac{\displaystyle a_{nm}}{\displaystyle c_{nm}}=%
-\frac{\displaystyle r_{\mathbf{v}+\mathbf{j}}}%
{\displaystyle r_{\mathbf{v}-\mathbf{k}}}\cdot%
\frac{\displaystyle \Theta(P_2^-,\mathbf{v}+\mathbf{j})%
\Theta(P_1^+,\mathbf{v}-\mathbf{i}+\mathbf{j})%
\Theta(P_3^+,\mathbf{v}-\mathbf{i}-\mathbf{k})}%
{\displaystyle \Theta(P_2^-,\mathbf{v}-\mathbf{i}+\mathbf{j})%
\Theta(P_1^+,\mathbf{v}-\mathbf{i}-\mathbf{k})%
\Theta(P_3^+,\mathbf{v}-\mathbf{k})}\times
$$
$$
\times\exp(\int\limits_{P_0}^{P_2^-}\Omega_1+%
\!\!\int\limits_{P_0}^{P_1^+}(\Omega_2+\Omega_3)-%
\!\!\int\limits_{P_0}^{P_3^+}\Omega_1),
$$
$$
\frac{\displaystyle b_{nm}}{\displaystyle c_{nm}}=%
-\frac{\displaystyle r_{\mathbf{v}+\mathbf{j}}}%
{\displaystyle r_{\mathbf{v}-\mathbf{i}+\mathbf{j}}}\cdot%
\frac{\displaystyle \Theta(P_2^-,\mathbf{v}+\mathbf{j})}%
{\displaystyle \Theta(P_2^-,\mathbf{v}-\mathbf{i}+\mathbf{j})}%
\cdot\exp(\int\limits_{P_0}^{P_2^-}\Omega_1),
$$
$$
\frac{\displaystyle d_{nm}}%
{\displaystyle c_{nm}}=%
\frac{\displaystyle r_{\mathbf{v}+\mathbf{j}}}%
{\displaystyle r_{\mathbf{v}-\mathbf{i}-\mathbf{k}}}\cdot%
\frac{\displaystyle \Theta(P_2^-,\mathbf{v}+\mathbf{j})%
\Theta(P_1^+,\mathbf{v}-\mathbf{i}+\mathbf{j})}%
{\displaystyle \Theta(P_2^-,\mathbf{v}-\mathbf{i}+\mathbf{j})%
\Theta(P_1^+,\mathbf{v}-\mathbf{i}-\mathbf{k})}\cdot%
\exp(\int\limits_{P_0}^{P_2^-}\Omega_1+%
\!\!\int\limits_{P_0}^{P_1^+}(\Omega_2+\Omega_3)),
$$
$$
\frac{\displaystyle v_{nm}}{\displaystyle c_{nm}}=%
\frac{\displaystyle r_{\mathbf{v}+\mathbf{j}}}%
{\displaystyle r_{\mathbf{v}}}\cdot%
\frac{\displaystyle \Theta(P_2^-,\mathbf{v}+\mathbf{j})%
\Theta(P_1^+,\mathbf{v}-\mathbf{i}+\mathbf{j})}%
{\displaystyle \Theta(P_2^-,\mathbf{v}-\mathbf{i}+\mathbf{j})%
\Theta(P_1^+,\mathbf{v}-\mathbf{i}-\mathbf{k})}\times
$$
$$
\times\exp(\int\limits_{P_0}^{P_2^-}\Omega_1+%
\!\!\int\limits_{P_0}^{P_1^+}(\Omega_2+\Omega_3)-%
\!\!\int\limits_{P_0}^{P_2+}\Omega_3)\cdot%
\biggl[\frac{\displaystyle \Theta(P_3^+,\mathbf{v}-\mathbf{i}-\mathbf{k})%
\Theta(P_2^+,\mathbf{v}-\mathbf{k})}%
{\displaystyle \Theta(P_3^+,\mathbf{v}-\mathbf{k})%
\Theta(P_2^+,\mathbf{v})}\times
$$
$$
\times\exp(-\!\!\int\limits_{P_0}^{P_3^+}\Omega_1)-
\frac{\displaystyle \Theta(P_2^+,\mathbf{v}-\mathbf{i}-\mathbf{k})}%
{\displaystyle \Theta(P_2^+,\mathbf{v})}%
\cdot\exp(-\!\!\int\limits_{P_0}^{P_2^+}\Omega_1)\biggr],
$$
where $\mathbf{v}=\mathbf{v}(n,m).$

\noindent{\sc Proof.}
Let $L\psi=0.$ Let us consider the case 
$n+m\equiv 0\,\!\!\! \pmod 2.$
Thus, the formula $L\psi=0$ becomes
\begin{equation}
a_{nm}\phi_{\mathbf{v}+\mathbf{i}-\mathbf{j}}+%
b_{nm}\phi_{\mathbf{v}+\mathbf{k}}+%
c_{nm}\phi_{\mathbf{v}+\mathbf{i}+\mathbf{k}}+
d_{nm}\phi_{\mathbf{v}-\mathbf{j}}+%
v_{nm}\phi_{\mathbf{v}}=0.
\label{lpnol}
\end{equation}
Consider the point $P^-_1.$ Let $\lambda$ be a local parameter
in a neighbourhood of $P_1^-.$ Hence
$\phi_{\alpha\beta\gamma}=\lambda^{-\alpha}\cdot h,$
where $h$ is a holomorphic 
function. The function $\exp\int^{P}_{P_0}\Omega_1$ has 
a pole of first order at
the point $P^-_1.$ Thus
$\exp\int^{P}_{P_0}\Omega_1=K_1^-\lambda^{-1}+\dots,$
where $K_1^-$ is a constant.
Therefore we have
$$
\phi_{\alpha\beta\gamma}(P)=r_{\alpha\beta\gamma}(K_1^-)^\alpha\!\!%
\left(\exp\int\limits_{P_0}^{P_1^-}\Omega_2\right)^{\beta}\!\!\!%
\left(\exp\int\limits_{P_0}^{P_1^-}\Omega_3\right)^{\gamma}\!\!\!%
\frac{\Theta(P_1^-,\alpha,\beta,\gamma)}{\Theta(P_1^-,0,0,0)}%
\lambda^{-\alpha}+\dots
$$
for $P$ in the neighbourhood of $P_1^-.$

Now we can write down the term with $\lambda^{-(\alpha(n,m)+1)}$
in the formula~(\ref{lpnol})
in the neighbourhood of $P_1^-$
$$
a_{nm}(K_1^-)^{\alpha(n,m)+1}r_{\mathbf{v}+\mathbf{i}-\mathbf{j}}%
\left(\exp\int\limits_{P_0}^{P_1^-}\Omega_2\right)^{\beta(n,m)-1}%
\left(\exp\int\limits_{P_0}^{P_1^-}\Omega_3\right)^{\gamma(n,m)}\times
$$
$$
\times\frac{\Theta(P_1^-,\mathbf{v}+\mathbf{i}-\mathbf{j})}%
{\Theta(P_1^-,0,0,0)}\lambda^{-(\alpha(n,m)+1)}+%
c_{nm}(K_1^-)^{\alpha(n,m)+1}r_{\mathbf{v}+\mathbf{i}+\mathbf{k}}\times
$$
$$
\times\left(\exp\int\limits_{P_0}^{P_1^-}\Omega_2\right)^{\beta(n,m)}%
\!\!\!\left(\exp\int\limits_{P_0}^{P_1^-}%
\Omega_3\right)^{\gamma(n,m)+1}%
\!\!\!\!\!\!\!\!\!\!%
\frac{\Theta(P_1^-,\mathbf{v}+\mathbf{i}+\mathbf{k})}%
{\Theta(P_1^-,0,0,0)}\lambda^{-(\alpha(n,m)+1)}=0.
$$
After simplifications we obtain a linear equation for the 
$a_{nm}$ and the $c_{nm}$
$$
a_{nm}r_{\mathbf{v}+\mathbf{i}-\mathbf{j}}%
\Theta(P_1^-,\mathbf{v}+\mathbf{i}-\mathbf{j})+%
c_{nm}r_{\mathbf{v}+\mathbf{i}+\mathbf{k}}%
\exp(\!\int\limits_{P_0}^{P_1^-}(\Omega_2+\Omega_3))%
\Theta(P_1^-,\mathbf{v}+\mathbf{i}+\mathbf{k})=0.
$$
Doing analogous computations at the points
$P_3^-,$ $P_2^-$ and $P_2^+$ we obtain
three other linear equations for 
the $a_{nm},$ $b_{nm},$ $c_{nm},$ $d_{nm}$ and $v_{nm}.$ 
These equations can be explicitely solved and
the formulas
for the coefficients of the operator $L$ given in the statement of 
the theorem are obtained.
The case $n+m\equiv 1\!\!\! \pmod 2$ is analogous.

Now let us suppose that the $a_{nm},\dots,v_{nm}$ are given by
the formulas of the statement of the theorem. Let us prove  
that $L\psi=0.$  Consider the case $n+m\equiv 0\pmod 2.$ 
Let us consider a function
$$
(\hat{L}\psi)_{nm}=\frac{a_{nm}}{d_{nm}}\psi_{n-1,m}+%
\frac{b_{nm}}{d_{nm}}\psi_{n+1,m}+%
\frac{c_{nm}}{d_{nm}}\psi_{n,m-1}+%
\frac{v_{nm}}{d_{nm}}\psi_{n,m}.
$$
This function has the same pole or zero structure
as the function $-\psi_{n,m+1}$ at the points
$P_i^{+},i=1,2,3.$
It follows from the formulas 
in the statement of the theorem
that $(\hat{L}\psi)_{nm}$ and $-\psi_{n,m+1}$
have the same pole or zero structure at the points
$P_i^{-},i=1,2,3.$
If a point $P\in\Gamma\setminus\{P_1^{\pm},P_2^{\pm},P_3^{\pm}\}$
is a pole of $(\hat{L}\psi)_{nm}$ or $-\psi_{n,m+1},$
then $P\in\mathcal{D}.$
Thus, by the Lemma,
$(\hat{L}\psi)_{nm}$ and $-\psi_{n,m+1}$ are proportional.
Moreover, from the formulas for the coefficients of
the operator $L$ it follows that the
terms with $\lambda^{\beta(n,m)-1}$ in the
series expansions of these two functions
at the point $P_2^+$ are the same.
Hence
$(\hat{L}\psi)_{nm}=-\psi_{n,m+1},$ but this is equivalent to
$L\psi=0.$ 
The case $n+m\equiv 1\!\!\! \pmod 2$ is analogous. 
This completes the proof. $\square$.

Any set of non-zero constants $g_{nm}$ defines a ``gauge'' transformation
of operators of the form~(\ref{L}) such that
$$
a'_{nm}=g^{-1}_{n-1,m}a_{nm}, \quad%
b'_{nm}=g^{-1}_{n+1,m}b_{nm}, \quad%
c'_{nm}=g^{-1}_{n,m-1}c_{nm}, 
$$
$$
d'_{nm}=g^{-1}_{n,m+1}d_{nm}, \quad%
v'_{nm}=g^{-1}_{nm}v_{nm}.
$$
This gauge transform acts on the eigenfunctions in the
following manner:
$\psi'_{nm}=g_{nm}\psi_{nm}.$

The following theorem is an easy corollary of the Theorem~1.

\noindent {\sc Theorem~$1'$.} For any set of ``spectral data''
consisting of: a non-singular curve $\Gamma$ of genus
$g,$ six points $P_i^{\pm}\in\Gamma,i=1,2,3,$ and a 
generic divisor $\mathcal{D}$ of $g$ points 
different from the $P_i^{\pm},$
there exists, up to a gauge transformation,
a unique operator $L$ of the form~(\ref{L}).

\section{The hexagonal operators: Novikov's class}\label{hexagonal}

Consider a triangular lattice in a plane.
We will use as coordinates triples of integers $k,l,m$ such that
$k+l+m=0.$

Consider an arbitrary
two-dimensional difference operator $L$ of the form~(\ref{L1}).
Our goal is to find some solution of the generalized inverse problem
stated in the introduction.

Our construction is as follows. Let $\Gamma$ be a nonsingular curve of genus
$g.$ Let $Q_i,$ $R_i,$ $i=1,2,3,$
be six points on $\Gamma.$ 
Let $\mathcal D$ be a generic divisor of the form
$\mathcal{D}=P_1+\dots+P_g$ such that the points
$P_k$ are 
different from the $Q_i,$ $R_i.$
Consider a function
$\phi_{\alpha\beta\gamma\rho\sigma\tau},$
$\alpha,$ $\beta,$ $\gamma,$ $\rho,$ $\sigma,$ $\tau\in\mathbb{Z},$
$\alpha+\beta+\gamma=0,$ $\rho+\sigma+\tau=0,$
defined on $\Gamma$ such that:

\noindent 1) If a point
$P\in\Gamma\setminus\{Q_1,Q_2,Q_3,R_1,R_2,R_3\}$
is a pole of $\phi_{\alpha\beta\gamma\rho\sigma\tau},$
then $P$ is one of the points
$P_k;$

\noindent 2) The function $\phi_{\alpha\beta\gamma\rho\sigma\tau}$ has
a pole of $\alpha$th order in $Q_1,$
a pole of $\beta$th order in $Q_2,$
a pole of $\gamma$th order in $Q_3;$
the same structure for $\rho,$ $\sigma,$ $\tau$ and
$R_1,$ $R_2,$ $R_3.$

\noindent{\sc Lemma} 1) Such a function
$\phi_{\alpha\beta\gamma\rho\sigma\tau}$ exists and
is unique up to multiplication by a constant.

\noindent 2) The explicit formula for
$\phi_{\alpha\beta\gamma\rho\sigma\tau}$ is:
$$
r_{\alpha\beta\gamma\rho\sigma\tau}\cdot%
\exp\int_{P_0}^{P}(\alpha\Omega_{Q_3Q_1}+\beta\Omega_{Q_3Q_2}+%
\rho\Omega_{R_3R_1}+\sigma\Omega_{R_3R_2})\times
$$
$$
\times\frac{\Theta(A(P)+\alpha U_{Q_3Q_1}+\beta U_{Q_3Q_2}+%
\rho U_{R_3R_1}+\sigma U_{R_3R_2}-A(\mathcal{D})-\mathcal{K})}%
{\Theta(A(P)-A(\mathcal{D})-\mathcal{K})},
$$
where $r_{\alpha\beta\gamma\rho\sigma\tau}$ is an arbitrary constant,
$P_0$ is
a fixed point defining the Abel transform $A$ (it should be remarked
that the paths of integration in $\int_{P_0}^P$ and in the Abel transform
are the same).

\noindent{\sc Proof} is done by a standard reasoning of the
theory of the algebro-geometric integration. $\square$

As in section~\ref{cross} we will use a vectorial notation.
We will represent the six integer numbers $\alpha,$ $\beta,$
$\gamma,$ $\rho,$ $\sigma,$ $\tau$ as one vector
$$
\mathbf{v}=\alpha\mathbf{e}_1+\beta\mathbf{e}_2+\gamma\mathbf{e}_3+%
\rho\mathbf{e}_4+\sigma\mathbf{e}_5+\tau\mathbf{e}_6\in\mathbb{Z}^6.
$$
Thus, we will write $\phi_{\mathbf{v}}$ instead of
$\phi_{\alpha\beta\gamma\rho\sigma\tau}.$

The key idea of the construction of our functions
$\psi_{klm}$ is a convinient relabelling in following way:
$\psi_{klm}=\phi_{\mathbf{v}(k,l,m)},$
where
$$
\left\{%
\begin{array}{l}
\mathbf{v}(k,l,m)=%
\frac{k-l}{3}\mathbf{e}_1+\frac{l-m}{3}\mathbf{e}_2+%
\frac{m-k}{3}\mathbf{e}_3+\frac{k-l}{3}\mathbf{e}_4+%
\frac{l-m}{3}\mathbf{e}_5+\frac{m-k}{3}\mathbf{e}_6\\
\mbox{if}\quad k-l=0 \pmod 3,\\
\mathbf{v}(k,l,m)=%
\frac{k-l-1}{3}\mathbf{e}_1+\frac{l-m+2}{3}\mathbf{e}_2+%
\frac{m-k-1}{3}\mathbf{e}_3+\frac{k-l+2}{3}\mathbf{e}_4+%
\frac{l-m-1}{3}\mathbf{e}_5+\\
+\frac{m-k-1}{3}\mathbf{e}_6 \quad \mbox{if}\quad k-l=1 \pmod 3,\\
\mathbf{v}(k,l,m)=%
\frac{k-l+1}{3}\mathbf{e}_1+\frac{l-m+1}{3}\mathbf{e}_2+%
\frac{m-k-2}{3}\mathbf{e}_3+\frac{k-l+1}{3}\mathbf{e}_4+%
\frac{l-m-2}{3}\mathbf{e}_5+\\
+\frac{m-k+1}{3}\mathbf{e}_6 \quad \mbox{if}\quad  k-l=2 \pmod 3.
\end{array}
\right.
$$

We will also use the following notation
$$
\Theta(P,\alpha\mathbf{e}_1+\beta\mathbf{e}_2+\gamma\mathbf{e}_3+%
\rho\mathbf{e}_4+\sigma\mathbf{e}_5+\tau\mathbf{e}_6)%
=\Theta(P,\alpha,\beta,\gamma,\rho,\sigma,\tau)=
$$
$$
=\Theta(A(P)+\alpha U_{Q_3Q_1}+\beta U_{Q_3Q_2}+%
\rho U_{R_3R_1}+\sigma U_{R_3R_2}-A(\mathcal{D})-\mathcal{K}).
$$

Let us formulate our theorem.

\noindent{\sc Theorem 2.} Let a family $\psi_{klm}$ be defined
as stated above. Then
$L\psi=0$ if and only if 
the coeficients
$a_{klm}$, $b_{klm}$, $c_{klm}$, $d_{klm}$,
$f_{klm},$ $g_{klm},$ of the operator $L$
are defined up to a multiplication by a constant by the following
formulas:

\noindent 1) if $k-l\equiv 0 \pmod 3,$ then
$$
\frac{\displaystyle a_{klm}}{\displaystyle b_{klm}}=%
\frac{\displaystyle r_{\mathbf{v}+\mathbf{e}_4-\mathbf{e}_5}}%
{\displaystyle r_{\mathbf{v}+\mathbf{e}_2-\mathbf{e}_3}}%
\cdot\biggl[\frac{\displaystyle\Theta(Q_2,%
\mathbf{v}-\mathbf{e}_1+\mathbf{e}_2)%
\Theta(Q_3,\mathbf{v}+\mathbf{e}_4-\mathbf{e}_5)}%
{\displaystyle\Theta(Q_2,%
\mathbf{v}+\mathbf{e}_2-\mathbf{e}_3)%
\Theta(Q_3,\mathbf{v}-\mathbf{e}_1+\mathbf{e}_2)}\times%
$$
$$
\times\exp(\int\limits_{P_0}^{Q_3}(\Omega_{R_2R_1}-\Omega_{Q_1Q_2})-%
\int\limits_{P_0}^{Q_2}\Omega_{Q_3Q_1})+
$$
$$
+\frac{\displaystyle\Theta(Q_2,%
\mathbf{v}+\mathbf{e}_2-\mathbf{e}_3+\mathbf{e}_4-\mathbf{e}_6)%
\Theta(R_1,\mathbf{v}+\mathbf{e}_4-\mathbf{e}_5)}%
{\displaystyle\Theta(Q_2,\mathbf{v}+\mathbf{e}_2-\mathbf{e}_3)%
\Theta(R_1,\mathbf{v}+\mathbf{e}_2-\mathbf{e}_3+\mathbf{e}_4-\mathbf{e}_6)}%
\times
$$
$$
\times%
\exp(\int\limits_{P_0}^{Q_2}\Omega_{R_3R_1}-%
\int\limits_{P_0}^{R_1}(\Omega_{Q_3Q_2}+\Omega_{R_3R_2}))%
\biggr],
$$
$$
\frac{\displaystyle d_{klm}}{\displaystyle b_{klm}}=%
-\frac{\displaystyle r_{\mathbf{v}+\mathbf{e}_4-\mathbf{e}_5}}%
{\displaystyle r_{\mathbf{v}-\mathbf{e}_1+\mathbf{e}_2}}\cdot%
\frac{\displaystyle\Theta(Q_3,\mathbf{v}+\mathbf{e}_4-\mathbf{e}_5)}%
{\displaystyle\Theta(Q_3,\mathbf{v}-\mathbf{e}_1+\mathbf{e}_2)}\cdot%
\exp(\int\limits_{P_0}^{Q_3}(\Omega_{R_2R_1}-\Omega_{Q_1Q_2})),
$$
$$
\frac{\displaystyle f_{klm}}{\displaystyle b_{klm}}=%
-\frac{\displaystyle r_{\mathbf{v}+\mathbf{e}_4-\mathbf{e}_5}}%
{\displaystyle r_{%
\mathbf{v}+\mathbf{e}_2-\mathbf{e}_3+\mathbf{e}_4-\mathbf{e}_6}}\cdot%
\frac{\displaystyle\Theta(R_1,\mathbf{v}+\mathbf{e}_4-\mathbf{e}_5)}%
{\displaystyle\Theta(Q_3,%
\mathbf{v}+\mathbf{e}_2-\mathbf{e}_3+\mathbf{e}_4-\mathbf{e}_6)}\times
$$
$$
\times\exp(\int\limits_{P_0}^{R_1}(-\Omega_{Q_3Q_2}-\Omega_{R_3R_2})),\quad
\frac{\displaystyle c_{klm}}{\displaystyle b_{klm}}=0,\quad
\frac{\displaystyle g_{klm}}{\displaystyle b_{klm}}=0,
$$
where $\mathbf{v}=\mathbf{v}(k,l,m),$

\noindent 2) if $k-l\equiv 1 \pmod 3,$ then
$$
\frac{\displaystyle b_{klm}}{\displaystyle d_{klm}}=%
-\frac{\displaystyle r_{\mathbf{v}-\mathbf{e}_4+\mathbf{e}_6}}%
{\displaystyle r_{\mathbf{v}+\mathbf{e}_1-\mathbf{e}_2-%
\mathbf{e}_5+\mathbf{e}_6}}\cdot%
\frac{\displaystyle\Theta(R_3,\mathbf{v}-\mathbf{e}_4+\mathbf{e}_6)}%
{\displaystyle\Theta(R_3,\mathbf{v}+\mathbf{e}_1-\mathbf{e}_2-%
\mathbf{e}_5+\mathbf{e}_6)}\times
$$
$$
\times\exp(\int\limits_{P_0}^{R_3}(\Omega_{Q_1Q_2}+\Omega_{R_1R_2})),
$$
$$
\frac{\displaystyle c_{klm}}{\displaystyle d_{klm}}=%
\frac{\displaystyle r_{\mathbf{v}-\mathbf{e}_4+\mathbf{e}_6}}%
{\displaystyle r_{\mathbf{v}+\mathbf{e}_1-\mathbf{e}_2}}\cdot%
\biggl[\frac{\displaystyle\Theta(R_1,\mathbf{v}+%
\mathbf{e}_1-\mathbf{e}_2-\mathbf{e}_5+\mathbf{e}_6)%
\Theta(R_3,\mathbf{v}-\mathbf{e}_4+\mathbf{e}_6)}%
{\displaystyle\Theta(R_1,\mathbf{v}+\mathbf{e}_1-\mathbf{e}_2)%
\Theta(R_3,\mathbf{v}+\mathbf{e}_1-%
\mathbf{e}_2-\mathbf{e}_5+\mathbf{e}_6)}\times
$$
$$
\times\exp(\int\limits_{P_0}^{R_3}%
(\Omega_{Q_1Q_2}+\Omega_{R_1R_2})-%
\int\limits_{P_0}^{R_1}\Omega_{R_3R_2})+
$$
$$
+\frac{\displaystyle\Theta(R_1,\mathbf{v}+\mathbf{e}_1-\mathbf{e}_3)%
\Theta(Q_2,\mathbf{v}-\mathbf{e}_4+\mathbf{e}_6)}%
{\displaystyle\Theta(R_1,\mathbf{v}+\mathbf{e}_1-\mathbf{e}_2)%
\Theta(Q_2,\mathbf{v}+\mathbf{e}_1-\mathbf{e}_3)}\times
$$
$$
\times\exp(\int\limits_{P_0}^{R_1}\Omega_{Q_3Q_2}-%
\int\limits_{P_0}^{Q_2}(\Omega_{Q_3Q_1}+\Omega_{R_3R_1}))\biggr],
$$
$$
\frac{\displaystyle f_{klm}}{\displaystyle d_{klm}}=%
-\frac{\displaystyle r_{\mathbf{v}-\mathbf{e}_4+\mathbf{e}_6}}%
{\displaystyle r_{\mathbf{v}+\mathbf{e}_1-\mathbf{e}_3}}\cdot%
\frac{\displaystyle\Theta(Q_2,\mathbf{v}-\mathbf{e}_4+\mathbf{e}_6)}%
{\displaystyle\Theta(Q_2,\mathbf{v}+\mathbf{e}_1-\mathbf{e}_3)}\cdot%
\exp(\int\limits_{P_0}^{Q_2}(-\Omega_{Q_3Q_1}-\Omega_{R_3R_1})),
$$
$$
\frac{\displaystyle a_{klm}}{\displaystyle d_{klm}}=0,\quad%
\frac{\displaystyle g_{klm}}{\displaystyle d_{klm}}=0,
$$
where $\mathbf{v}=\mathbf{v}(k,l,m),$

\noindent 3) if $k-l\equiv 2 \pmod 3,$ then
$$
\frac{\displaystyle b_{klm}}{\displaystyle f_{klm}}=%
-\frac{\displaystyle r_{\mathbf{v}+\mathbf{e}_5-\mathbf{e}_6}}%
{\displaystyle r_{\mathbf{v}-\mathbf{e}_2+\mathbf{e}_3}}\cdot%
\frac{\displaystyle\Theta(Q_1,\mathbf{v}+\mathbf{e}_5-\mathbf{e}_6)}%
{\displaystyle\Theta(Q_1,\mathbf{v}-\mathbf{e}_2+\mathbf{e}_3)}\cdot%
\exp(\int\limits_{P_0}^{Q_1}(\Omega_{Q_3Q_2}+\Omega_{R_3R_2})),
$$
$$
\frac{\displaystyle d_{klm}}{\displaystyle f_{klm}}=%
-\frac{\displaystyle r_{\mathbf{v}+\mathbf{e}_5-\mathbf{e}_6}}%
{\displaystyle r_{\mathbf{v}-\mathbf{e}_1+\mathbf{e}_3-%
\mathbf{e}_4+\mathbf{e}_5}}\cdot%
\frac{\displaystyle\Theta(R_2,\mathbf{v}+\mathbf{e}_5-\mathbf{e}_6)}%
{\displaystyle\Theta(R_2,\mathbf{v}-\mathbf{e}_1+\mathbf{e}_3-%
\mathbf{e}_4+\mathbf{e}_5)}\times
$$
$$
\times\exp(\int\limits_{P_0}^{R_2}(\Omega_{Q_3Q_1}+\Omega_{R_3R_1})),
$$
$$
\frac{\displaystyle g_{klm}}{\displaystyle f_{klm}}=%
\frac{\displaystyle r_{\mathbf{v}+\mathbf{e}_5-\mathbf{e}_6}}%
{\displaystyle r_{\mathbf{v}-\mathbf{e}_1+\mathbf{e}_3}}\cdot%
\left[\frac{\displaystyle\Theta(Q_3,\mathbf{v}-\mathbf{e}_2+\mathbf{e}_3)%
\Theta(Q_1,\mathbf{v}+\mathbf{e}_5-\mathbf{e}_6)}%
{\displaystyle\Theta(Q_3,\mathbf{v}-\mathbf{e}_1+\mathbf{e}_3)%
\Theta(Q_1,\mathbf{v}-\mathbf{e}_2+\mathbf{e}_3)}\times\right.
$$
$$
\times\exp(\int\limits_{P_0}^{Q_1}(\Omega_{Q_3Q_2}+\Omega_{R_3R_2})-%
\int\limits_{P_0}^{Q_3}\Omega_{Q_1Q_2})+
$$
$$
+\frac{\displaystyle\Theta(Q_3,\mathbf{v}-%
\mathbf{e}_1+\mathbf{e}_3-\mathbf{e}_4+\mathbf{e}_5)%
\Theta(R_2,\mathbf{v}+\mathbf{e}_5-\mathbf{e}_6)}%
{\displaystyle\Theta(Q_3,\mathbf{v}-\mathbf{e}_1+\mathbf{e}_3)%
\Theta(R_2,\mathbf{v}-\mathbf{e}_1+\mathbf{e}_3%
-\mathbf{e}_4+\mathbf{e}_5)}\times
$$
$$
\left.\times\exp(\int\limits_{P_0}^{R_2}(\Omega_{Q_3Q_1}+\Omega_{R_3R_1})%
+\int\limits_{P_0}^{Q_3}\Omega_{R_1R_2})\right],
$$
$$
\frac{\displaystyle a_{klm}}{\displaystyle f_{klm}}=0,\quad
\frac{\displaystyle c_{klm}}{\displaystyle f_{klm}}=0,
$$
where $\mathbf{v}=\mathbf{v}(k,l,m).$

\noindent{\sc Proof.} The proof is analogous to the proof
of the Theorem~1. $\square$

Any set of non-zero constants $h_{klm}$ defines a ``gauge'' transformation
of operators of the form~(\ref{L1}) such that
$$
a'_{klm}=h^{-1}_{k,l+1,m-1}a_{klm}, \quad%
b'_{klm}=h^{-1}_{k,l-1,m+1}b_{klm}, \quad%
c'_{klm}=h^{-1}_{k+1,l-1,m}c_{klm}, 
$$
$$
d'_{klm}=h^{-1}_{k-1,l+1,m}d_{klm}, \quad%
f'_{klm}=h^{-1}_{k+1,l,m-1}f_{klm},\quad%
g'_{klm}=h^{-1}_{k-1,l,m+1}g_{klm}.
$$
This gauge transform acts on the eigenfunctions 
in the following manner:
$\psi'_{klm}=h_{klm}\psi_{klm}.$

The following theorem is an easy corollary of the Theorem~2.

\noindent {\sc Theorem~$2'$.} For any set of ``spectral data''
consisting of: a non-singular curve $\Gamma$ of genus
$g,$ six points $Q_i,R_i\in\Gamma,i=1,2,3,$ and a 
generic divisor $\mathcal{D}$ of $g$ points 
different from the $Q_i,R_i,$
there exists, up to a gauge transformation,
a unique operator $L$ of the form~(\ref{L1}).

\section*{Acknowledgements}

The authors are indebted to Professor Alexander P.~Veselov 
for suggesting this problem and fruitful discussions.
The authors also thank Professor Pavel Winternitz for
discussions. The authors also thank K.~Thomas for the help
in the preparation of the manuscript.
The authors wish to thank the referees for
useful remarks.

The main part of this research was performed 
during the participation of one of the authors (A.O.) in
the S\'eminaire de Math\'ematiques Sup\'erieures at
the Universit\'e de Montr\'eal in the summer of 1999 and he is
very grateful to the Universit\'e de Montr\'eal for
hospitality.

During this work the authors were supported
by the grant INTAS~96-0770 (A.O.)
and fellowships from the Institut de Sciences Math\'ematiques
and the Universit\'e de Montr\'eal (A.P.), which
are gratefully acknowledged.


\begin{thebibliography}{9}
\bibitem{DKN} B.~A.~Dubrovin, I.~M.~Krichever, S.~P.~Novikov.
The Scr\"odinger equation in a periodic field
and Riemann surfaces.
Dokl. Akad. Nauk SSSR, \textbf{229} (1976), no.~1.,
p.~15-18. (in Russian).
English translation: Soviet Math. Dokl. \textbf{17} (1976),
no.~4, p.~947-951.
\bibitem{NV} A.~P.~Veselov, S.~P.~Novikov. Finite-zone,
two-dimensional Schr\"odinger operators. Potential operators.
Dokl. Akad. Nauk SSSR, \textbf{279} (1984), no.~4, p.~784-788.
(in Russian). English translation: Soviet Math. Dokl.,
\textbf{30} (1984), no.~3, p.~705-708.
\bibitem{K} I.~M.~Krichever. Two-dimensional periodic
difference operators and algebraic geometry. 
Dokl. Akad. Nauk SSSR, \textbf{285} (1985), no.~1, p.~31-36. 
(in Russian).
English translation: Soviet Math. Dokl. \textbf{32} (1985),
no.~3, p.~623-627.
English translation: Russian Math. Surveys, \textbf{52} (1997),
no.~5, p.~1057-1116.
\bibitem{N} S.~P.~Novikov. Algebraic properties of two-dimensional
difference operators. 
Uspekhi. Mat. Nauk, \textbf{52} (1997), no.~1, p.~225-226.
(in Russian).
English translation: Russian Math. Surveys, \textbf{52} (1997),
no.~1, p.~226-227.
\bibitem{NVC} S.~P.~Novikov, A.~P.~Veselov. Exactly solvable 
two-dimensional Schr\"odinger operators and
Laplace transformations. In {\em Solitons, geometry, and topology: on
the crossroad,} Ed. V.~M.~Buchstaber and S.~P.~Novikov, AMS Transl.
Ser.~2, \textbf{179} (1997), p.~109-132. 
\bibitem{ND} S.~P.~Novikov, I.~A.~Dynnikov. Discrete spectral
symmetries of small-dimensional differential operators and
difference operators on regular lattices and
two-dimensional manifolds. Uspekhi Mat. Nauk, \textbf{52} (1997),
no.~5, p.~175-234. (in Russian).
English translation: Russian Math. Surveys, \textbf{52} (1997), 
p.~1057-1116.
\bibitem{D} B.~A.~Dubrovin. Theta-functions and nonlinear
equations. Uspekhi. Mat. Nauk, \textbf{36} (1981), no.~2, p.~11-80.
(in Russian).
English translation: Russian Math. Surveys, \textbf{36} (1981),
no.~2, p.~11-92.
\end{thebibliography}
\end{document}